\documentclass[12pt]{article}   
\def\preprint#1{\thispagestyle{empty}~\newline\vspace*{-22.65mm}
\begin{flushright}\begin{tabular}{l} #1 \end{tabular}
\end{flushright}\vspace{1cm}}
\catcode`\@=11 \@addtoreset{equation}{section}
\def\theequation{\arabic{section}.\arabic{equation}}\catcode`\@=11
\usepackage{latexsym}
\begin{document}

\preprint{gr-qc/9803035}
\renewcommand{\theequation}{\arabic{section}.\arabic{equation}}

\begin{center}
{\large \bf Wormhole effective interactions in anti-de Sitter spacetime}\\[5mm]
{Carlos Barcel\'o$^{1,2}$ and Luis J. Garay$^2$}\\[3mm]
{\it $^1$ Instituto de Astrof\'{\i}sica de 
Andaluc\'{\i}a, CSIC\\ 
Camino Bajo de Hu\'etor, 18080 Granada, Spain\\
$^2$ Instituto de Matem\'aticas y 
F\'{\i}sica Fundamental, CSIC\\
C/  Serrano 121, 28006 Madrid, Spain}\\
10 March 1998
\end{center}
\begin{abstract}

The effects of asymptotically anti-de Sitter wormholes
in low-energy field theory  are calculated in full detail 
for three different matter contents: a conformal scalar field, 
an electromagnetic field and gravitons. 
There exists a close relation between the choice of vacuum 
for the matter fields and the selection of a basis of the
Hilbert space of anti-de Sitter wormholes. In the presence of 
conformal matter (i.e., conformal scalar or electromagnetic fields), 
this relation allows us to interpret  the elements of these bases 
as wormhole states containing a given number of particles.
This interpretation is subject to the same kind of ambiguity in
the definition of particle as that arising from quantum field
theory in curved spacetime. In the case of gravitons,
owing to the non-conformal coupling, it is not possible to describe 
wormhole states in terms of their particle content. 

\vskip 3mm
\noindent
{PACS: 04.60.Ds, 04.62.+v, 98.80 Hw}
\end{abstract}

\section{Introduction}

\setcounter{equation}{0}

Wormholes describe processes that involve baby
universes branching off and joining onto asymptotically large
regions of spacetime and have been extensively studied in
the literature \cite{colec1}. Their effects in low-energy fields
can be taken into account, in the
dilute wormhole approximation, by introducing local
effective interactions in a fixed background 
\cite{col,haw1}. As a consequence, wormholes
would modify the coupling constants of any low-energy effective
physical theory. These effective interactions produced by
wormholes have been calculated for a variety of matter fields
coupled to gravity \cite{haw,lyo,dow,lafl,gar,luk}. In all these
calculations, the asymptotically large region of the wormhole has
been assumed to be flat. Therefore, the effective interactions
produced by such wormholes have been modelled by interacting
terms added to a free quantum field theory in flat spacetime.

In Refs. \cite{nos,bg98}, we dealt with wormholes with
other types of asymptotic behaviour; in particular, we studied
the case of wormholes in asymptotically anti-de Sitter (adS)
spacetimes. This exhausts all possible maximally symmetric
asymptotic behaviours. Also, adS wormholes could be regarded as
excited states in the sense that the cosmological constant
$\Lambda$ could be interpreted as a non-vanishing asymptotic
energy of the matter field. 

Wormholes in
flat spacetime do not contribute to the cosmological constant
nor to the gravitational coupling constant $G$ directly 
\cite{haw1,lafl}. 
In Ref. \cite{bg98}, we showed that this is also true in adS 
spacetime. Furthermore, the interaction Lagrangians for the 
lower inhomogeneous modes in adS spacetime do not have any 
contribution from the cosmological constant and therefore 
have the same form as in the flat case. However, the interaction 
Lagrangians for the higher inhomogeneous modes in adS spacetime 
and those in the flat case differ in terms that are proportional 
to powers of the cosmological constant. In the case of gravitons, the
non-conformal nature of their coupling, although
present, is hidden in the flat case but explicit in adS
spacetimes. This introduces new elements in the analysis of the
latter case, such as the non-separability between the homogeneous
and inhomogeneous modes.
Another interesting characteristic of this model is that there
naturally exist different possible vacuum choices for the matter
fields in the asymptotic regions, as happens in quantum field
theory in curved spacetime. We saw in Ref. \cite{bg98}, and will be 
further discussed here,  that there exists a relation
between the definition of this vacuum and the choice of basis in
the Hilbert space of quantum wormholes, which allows us, for
conformally invariant matter, to
interpret the basis elements as states of wormholes containing
certain number of particles.

In this paper, we will consider three separate kinds of matter
fields coupled to gravity in the presence of a negative
cosmological constant, namely, a conformal scalar field, an
electromagnetic field and gravitons. 
We will model the effects of closed universes
that branch off or join onto an asymptotically adS spacetime 
in the dilute wormhole approximation by
means of an effective quantum field theory in an adS background 
\cite{col,banks}. In order to
find these effects on the matter fields, and following previous
analyses \cite{haw,lyo,dow,lafl,gar,luk,bg98}, we
will explicitly calculate matrix elements 
$\langle 0|\Phi(x_1)\cdots
\Phi(x_r)|\Psi_\alpha\rangle$ of products of operators
$\Phi(x)$, each representing a generic matter field at a
different point, between an specific vacuum $|0\rangle$ and the
elements $\{ \Psi_\alpha \}$ of a basis of the Hilbert space of
adS wormholes \cite{haw}. In flat spacetime, the state that is void of particles for
inertial observers defines a preferred vacuum for which
its associated propagator is asymptotically damped.
In adS spacetime, we can define a one-parameter family of 
maximally symmetric vacua by analogy with the de-Sitter
case\footnote{ There are some subtleties in the definition of 
states in an adS Lorentzian background because this spacetime 
does not posses a well-posed Cauchy problem. However, we will 
work in the Euclidean sector where this issue does not 
arise \cite{avis}.} \cite{burges,allen3}. 
The propagator associated with each vacuum of this family
is damped when the geodesic distance between the two points
becomes large, but with a different fall-off for each vacuum
choice \cite{allen1}. This family of vacua is the
analogue in adS spacetime of the vacua defined by uniformly
accelerated reference frames in the flat case. The state 
$|0\rangle$ is any fixed vacuum of this family.  

In order to calculate the
local effective interaction terms produced by adS wormholes, we
need to obtain the quantum wormhole wave functions for these
systems. This will be done in Sec. \ref{wwf}. Section \ref{tpf}
is devoted to calculate semiclassically the two-point functions
for each matter content in the presence of a wormhole. Processes
involving more complicated matrix elements could be treated in a
similar way. We will find the effective interaction terms that
have to be added to the bare action in order to reproduce these
Green functions in Sec. \ref{ei}. Finally, we will summarise the
results of this work and conclude in Sec. \ref{sc}.

\section{Wormhole wave functions}
\setcounter{equation}{0}
\label{wwf}

In order to find the wormhole wave functions, we have to solve
the WDW equation ${\cal H}\Psi_{\alpha}=0$ and the momenta
constraint equations $ {\cal H}_i\Psi_{\alpha}=0$ with
appropriate boundary conditions. These quantum constraints are
associated with the invariance of general relativity under time
reparametrisations (modulo spatial diffeomorphisms) and spatial
diffeomorphisms, respectively. They are difficult to solve
exactly and, therefore, we will perform a perturbative treatment.
We will take a homogeneous and isotropic spacetime,
characterised by its scale factor, which will be restricted to be
positive in order to avoid repetition of geometries, and the
homogeneous mode of the matter field as the configurations
variables that will be treated exactly. Then, we will decompose
the perturbation around the homogeneous configuration in
hypersherical harmonics on the three-sphere $S^3$. The
coefficients of this expansion will represent the rest of the
degrees of freedom.

\subsection{Conformal scalar field}

Let us now consider a scalar field conformally coupled to
gravity. Perturbatively, we can express the scalar field $\phi$
and the spatial part of the metric $g_{ij}$ as\footnote
{Throughout this paper, we will use a rescaled cosmological
constant $\lambda=-\frac{1}{3}\Lambda$ and set $\frac{3 }{2} \pi
m_p^2=1$.} 
\begin{equation}
\phi=\sqrt{\frac{1}{2 \pi^2}} a^{-1}
\sum_{n,\sigma_n}\chi_{n\sigma_n} Q^{n\sigma_n},
\label{dsc}
\end{equation}
\begin{equation}
g_{ij}= a^2 \Omega_{ij}+h_{ij},
\end{equation}
\begin{equation}
h_{ij}=\sum_{n,\sigma_n}\left(\sqrt{6}a_{n\sigma_n}
Q_{ij}^{n\sigma_n} + \sqrt{6}b_{n\sigma_n}
P_{ij}^{n\sigma_n} + \sqrt{2}c_{n\sigma_n} S_{ij}^{n\sigma_n} +
2d_{n\sigma_n} G_{ij}^{n\sigma_n}\right),  
\label{pmet}
\end{equation}
$\Omega_{ij}$ being the metric in the unit three-sphere. The
hyperspherical harmonics are defined as follows \cite{lifs}.
$Q^{n\sigma_n}$, $S_i^{n\sigma_n}$ and $ G_{ij}^{n\sigma_n}$ are
the scalar, transverse vector ($S_i^{n\sigma_n |i} =0 $) and
transverse traceless tensor ($G_{ij}^{n\sigma_n|j} =0$, $
G_i^{n\sigma_n i} =0$) harmonics. They are eigenfunctions of the
Laplace-Beltrami operator, $\Omega^{ij} \nabla_i \nabla_j$, in
$S^3$ with eigenvalues $-(n^2-1)$, $-(n^2-2)$ and $-(n^2-3)$,
respectively. The index $ \sigma_n$ runs over a basis of the
corresponding degenerate eigenspace. The remaining tensor
harmonics are defined as: 
\begin{equation}
Q_{ij}^{n\sigma_n} = \Omega_{ij}\frac{1}{3} Q^{n\sigma_n},
\hspace{5mm}
P_{ij}^{n\sigma_n} =\frac{1}{(n^2-1)} Q_{|ij}^{n\sigma_n} +
Q_{ij}^{n\sigma_n} ,\hspace{5mm} S_{ij}^{n\sigma_n} =
S^{n\sigma_n}_{(i|j)}.
\end{equation}

Among the degrees of freedom that we are considering, the
coefficients $a_{n\sigma_n}$, $b_{n\sigma_n}$ and $c_{n\sigma_n}$
are gauge; we can set them to zero by means of an appropriate
diffeomorphism in $S^3$ and introducing suitable lapse and shift
functions. The variables $d_{n\sigma_n}$, however, are true
degrees of freedom associated with gravitational waves. For the
time being, we set them to zero, $d_{n\sigma_n}=0$.

Then, the WDW equation becomes 
\begin{equation}
\left[-\frac{\partial^2}{\partial a^2}+a^2+\lambda a^4-
\sum_{n,\sigma_n} \left( -\frac{\partial^2}{\partial
\chi_{n\sigma_n}^2} +n^2\chi_{n \sigma_n}^2-\frac{1}{2}
\right)\right] \Psi(a,\chi_{n\sigma_n})=0.
\label{eqfsc}
\end{equation}
In the process of arriving at the WDW equation, we have chosen an
operator ordering that removes the ground-state energy of each
oscillator. Its solutions can be obtained by separation of
variables: 
\begin{equation}
\left[\sum_{n,\sigma_n} \left( -\frac{\partial^2} {\partial
\chi_{n\sigma_n}^2}+n^2\chi_{n\sigma_n}^2-\frac{1}{2}\right)\right]
\varphi_E (\chi_{n\sigma_n})=E\varphi_E (\chi_{n\sigma_n}),
\end{equation}
\begin{equation}
\left[-\frac{\partial^2}{\partial a^2}+a^2+ \lambda a^4
\right]\psi_E(a)=E\psi_E(a).
\end{equation}
We impose the standard boundary conditions for quantum harmonic
oscillators to the matter field part $\varphi_E $, so that $E$
will be a sum of harmonic oscillator energies. As for the
gravitational sector, we will only demand that $\psi_E $ is
exponentially damped when $a \rightarrow + \infty $. These
boundary conditions and their connection with the path integral
formalism were studied in detail in Ref. \cite{nos}. The wave
functions must be further restricted, owing to the linearisation
instability of the perturbations to the rotationally invariant
homogeneous configuration $(a, \chi_1)$ \cite{lafl,mon}. This
restriction should make the wave functions depend on the
inhomogeneous $n>1$ configuration variables only through the
rotationally invariant combination
$\chi_n^2=\sum_{\sigma_n}\chi_{n\sigma_n}^2 $.

We finally obtain wave functions of the form 
\begin{equation}
\Psi_{N_1,\cdots,N_n,\cdots } (a,\chi_n)= \psi_E(a)
{\cal H}_{N_1}(\chi_1)e^{-\frac{1}{2} \chi_1^2}
\prod_{n>1} {\cal L}_{N_n}^{(n^2-3)/2}(n\chi_n^2)
e^{-\frac{1}{2}n \chi_n^2}, 
\label{wdwsc}
\end{equation} 
where $E=N_1+\sum_{n>1} 2nN_n$, ${\cal H}_{N_1}$ is the Hermite
polynomial of degree $N_1$ and ${\cal L}_{N_n}^{(n^2-3)/2}$ is
the generalised Laguerre polynomial of degree $N_n$ in its
arguments \cite{abra}. These wave functions form an orthonormal
basis of the Hilbert space of wormhole quantum states with the
measure $\int da\prod_{n,\sigma_n} d\chi_{n\sigma_n}$, which is
proportional to
\begin{equation}
\int_0^{+\infty} da \int_{-\infty}^{+\infty}d\chi_1\prod_{n>1}
\int_0^{+\infty} d\chi_n (\chi_n)^{n^2-1},
\end{equation}
owing to the rotational invariance. The choice of this measure 
can be justified in the context of the algebraic quantization 
program (see Ref. \cite{nos}). Indeed, it is gauge dependent,
but this spurious dependence can be absorbed in the normalization
constants and therefore does not affect the results. Note that,
with this choice, the orthogonality relations are dictated by
the matter field part of the measure.

We can interpret $N_1$ as the
number of homogeneous scalar particles inside the wormhole. A
wave function with vanishig $N_n$'s except for a single mode
$n_0>1$ such that $N_{n_0}=1$, can be interpreted as a
rotationally invariant superposition of two-particle states in
the mode $n_0$. Then, $N_n$, with $n>1$, will be the number of
this kind of two-particle states inside the wormhole. As we will
see in Sec. \ref{tpf}, these particles must be associated with a
specific vacuum, which will be the one conformally related to the
natural vacuum for flat spacetimes.

\subsection{Electromagnetic field}

We can decompose the electromagnetic field $A_{\mu}$ in
harmonics in the three-sphere:
\begin{equation}
A_0=\sqrt{\frac{1}{2 \pi^2}}\sum_{n,\sigma_n}\alpha_{n\sigma_n}
Q^{n\sigma_n}, \hspace{10mm}
A_i=\sqrt{\frac{1}{2 \pi^2}}\sum_{n,\sigma_n}(\beta_{n\sigma_n}
S_i^{n\sigma_n}+\gamma_{n\sigma_n} P_i^{n\sigma_n}),
\end{equation}
where the harmonics $Q^{n\sigma_n}$, $S_i^{n\sigma_n}$ were
introduced in the previous subsection and the longitudinal
vector harmonics $P_i^{n\sigma_n}$ are defined through their
relation with the scalar ones $P_i^{n\sigma_n}=\frac{1}{n^2-1}
Q^{n\sigma_n}_{|i}$.

The wave functions do not depend on $\alpha_{n\sigma_n}$ and
$\gamma_{n\sigma_n}$ because these are Lagrange multipliers
\cite{dow}. They only depend on the transverse degrees of freedom
$\beta_{n\sigma_n}$. Therefore, the WDW equation for
electromagnetic wormholes in adS spacetime can be written as
\begin{equation}
\left[-\frac{\partial^2}{\partial a^2}+a^2+\lambda a^4 -
\sum_{n,\sigma_n} \left(-\frac{\partial^2}{\partial
\beta_{n\sigma_n}^2}+n^2\beta_{n
\sigma_n}^2-\frac{1}{2}\right)\right]
\Psi(a,\beta_{n\sigma_n})=0.
\label{wdwel}
\end{equation}
The transverse degrees of freedom can be separated into positive
and negative helicities because the helicity operator commutes
with rotations. Their respective coefficients in the harmonic
expansion will be called $\beta_{n\sigma_n}^{\pm} $. The
requirement that the wave functions must be invariant under
$SO(4)$ implies that they only depend on the invariant quantities
$(\beta_n^{\pm})^2=\sum_{\sigma_n}(\beta_{n \sigma_n}^{\pm})^2$,
where $\sigma_n$ runs over the corresponding positive and
negative helicity degenerate eigenspaces. As in the scalar field
case, an orthonormal basis of the Hilbert space of rotationally
invariant wave functions for this system is given by
\begin{equation}
\Psi_{N_2^{\pm}, \cdots ,N_n^{\pm}, \cdots} (a,\beta_n^{\pm})=
\psi_E(a) \prod_{n,p=\pm} {\cal
L}_{N_n^p}^{(n^2-4)/2}(n(\beta_n^p)^2) e^{-\frac{1}{2}n
(\beta_n^p)^2}, \label{emwf}
\end{equation}
where $E=\sum_{n=2} 2n(N_n^+ + N_n^-)$ and ${\cal
L}_{N_n^p}^{(n^2-4)/2}$ are the Laguerre polynomials. $N_n^{\pm}$
can be interpreted as the number of rotationally invariant
two-particle states of positive and negative helicity in the mode
$n$ inside the wormhole, which are associated with the conformal
vacuum as we will see in Sec \ref{tpf}. A wormhole cannot contain a
single-particle state because of the $SO(4)$ rotational
invariance. This symmetry forbids the existence of wormholes with
non-vanishing global spin.

\subsection{Gravitational perturbations}

The last case we are going to consider is that of gravitons,
represented by the coefficients $d_{n\sigma _n}$ of the
transverse traceless tensor harmonics $G_{ij}^{n\sigma _n}$ in
the expansion of the metric (\ref {pmet}). We expand the
constraints up to second order in perturbation theory and, thus,
the dynamical equations for $d_{n\sigma _n}$, obtained by varying
these constraints, will be first order in $d_{n\sigma _n}$. It
has been shown \cite{gris} that, when the Einstein tensor of the
background spacetime is traceless, the behaviour of
gravitational waves in that background is similar to that of a
conformal scalar field. For this reason, the WDW equation in the
flat case splits into two separate parts \cite{lafl}. However, if
the cosmological constant does not vanish, the trace of the
Einstein tensor is $-6\lambda $ and, therefore, the above
argument does not apply. In fact, we can write the WDW equation
for a non-vanishing cosmological constant as
\begin{equation}
\left[ -\frac{\partial ^2}{\partial {\tilde a}^2}+{\tilde
a}^2+\lambda { \tilde a}^4-\sum_{n,\sigma _n}\left(
-\frac{\partial ^2}{\partial {\tilde d} _{n\sigma
_n}^2}+n^2{\tilde d}_{n\sigma _n}^2+2\lambda {\tilde a}^2{\tilde
d}_{n\sigma _n}^2\right) \right] \Psi ({\tilde a},{\tilde
d}_{n\sigma _n})=0,
\label{eqfgr}
\end{equation}
where ${\tilde a}=a(1-3/2\sum d_{n\sigma _n}^2)$, ${\tilde
d}_{n\sigma _n}={\tilde a}d_{n\sigma _n}$ and we have redefined
the lapse function as ${\tilde N}=N(1+2\sum d_{n\sigma _n}^2)$.
In this expression, it is understood that the vacuum energies
have to be removed. 

The WDW equation is quite involved and we have not succeeded in
finding explicit solutions. However, when the scale factor
becomes large, the wave function will be picked around an
asymptotically adS spacetime and we will then recover a quantum
field theory of gravitons on the adS background (in this
approximation, the back-reaction can be neglected)
\cite{hall,wada}. Then, the semiclassical wave functions for the
graviton part will be harmonic oscillator eigenfunctions with a
frequency that depends on the scale factor \cite{wada}. These
considerations allow us to construct a formal basis of the
Hilbert space of this system, denoted by $\{\Psi_{\alpha}(a,
d_n^{\pm})\}$, in which the requirement of rotational invariance
has already been imposed.

\section{Two-point functions}

\setcounter{equation}{0}
\label{tpf}

We are interested in processes in which there is creation or
annihilation of particles associated with a particular vacuum in
two different points of a spacetime region where a wormhole end
is inserted. Following Hawking \cite{haw}, we first have to
calculate a path integral over geometries and matter fields that
induce a metric $g_{ij}^0$ and a value $\Phi^0$ for the matter
field in a cross section of the wormhole. Also, they must satisfy
some asymptotic requirements that guarantee the asymptotically
adS behaviour and that select a vacuum for the matter field in
the semiclassical approximation. After that, we will integrate
over all possible configurations $(g_{ij}^0$, $\Phi^0)$ weighted
with the wormhole wave function, i.e. 
\begin{equation}
\langle 0|\Phi(x_1) \Phi(x_2)|\Psi_\alpha\rangle = 
\int {\cal D}\Phi^0 {\cal D}g_{ij}^0 
\Psi_\alpha [g^{0}_{ij},\Phi^0] \int {\cal D}\Phi {\cal D}
g_{\mu \nu} \Phi(x_1) \Phi(x_2) e^{-I[g_{\mu \nu},\Phi]}.  
\label{apg}
\end{equation}
We will evaluate this path integral semiclassically. The action $
I[g_{\mu \nu},\Phi]$ contains a surface term that renders it
finite for classical solutions \cite{nos}, a necessary
requirement for the semiclassical approximation to be meaningful.
The saddle point solution for the gravitational part must be an
asymptotically adS spacetime. As far as the low-energy regime is
concerned, it can be taken to be pure adS spacetime outside a
three-sphere in which the wave function takes its arguments.

The position of the spacetime points $x_1$, $x_2$ can only be
specified modulo the isometries of Euclidean adS spacetime,
$SO(4,1)$. If $\Phi(x)$ is a saddle point solution for the matter
field, then, $\Phi_g(x)$, the transform of $\Phi$ under an
element $g^{-1}$ of the group of isometries $SO(4,1)$, is also a
solution. It has the form
\begin{equation}
\Phi_g(x)=T(x,x_{g})\Phi(x_{g}),
\label{t}
\end{equation}
where $x_{g}$ stands for the transform of $x$ under $g$ and
$T(x,x_{g})$ is the usual tensor transformation given by a
product of factors $\frac{\partial x_{g}}{\partial x}$, one for
each spacetime index of the matter field $\Phi$. As a
consequence, one has to average over this group. AdS spacetime is
isomorphic to the coset space $SO(4,1)/SO(4)$. A property of
unimodular groups \cite{bar} tells us that integrating a function
$F$ in the group $SO(4,1)$, using the invariant Haar measure in
that group, is equivalent to integrating first over the group
$SO(4)$ and, then, over the coset space, i.e. over adS
spacetime 
\begin{equation}
\int_{SO(4,1)} dg F(g) = \int_{\rm adS} d^4x_0 \sqrt{g(x_0)} 
\int_{SO(4)} dh F(x_0 h),
\end{equation}
$h$ being a generic element of the isotropy group $SO(4)$. We can
interpret this integral as an average over the orientations of
the wormhole and the positions in which it can be inserted. If we
expand the field in hyperspherical harmonics, the average over
orientations must be performed independently for each mode. In
flat spacetime, we obtain similar results if we use its own group
of isometries, the Euclidean group in four dimensions $E_4$
\cite{haw}.

Throughout this paper, we shall describe adS spacetime with two
different sets of coordinates: polar coordinates, $\{
\mu,\theta_i \}$, with a line element 
\begin{equation}
ds^2= d \mu^2 + C(\mu)^{-2} d\Omega_3^2(\theta_i),
\label{pmetric}
\end{equation}
$d\Omega_3^2(\theta_i)$ being the metric on the unit three-sphere and 
$C(\mu)=\frac{\sqrt{\lambda}}{\sinh (\sqrt{\lambda}\mu )}$,
and quasi-Euclidean coordinates, $\vec {x}=\{x^1,x^2,x^3,x^4\}$,
for which
\begin{equation}
ds^2=d{\vec {x}}^2- \frac{\lambda \left( \vec {x} \cdot 
d\vec{x}\right)^2}{(1+\lambda {\vec {x}}^2 )},  
\label{metric}
\end{equation}
where the dot represents Euclidean contraction of indices. The
change of coordinates that relates these two sets is of the form 
$x^a=C(\mu)^{-1}\widehat x^a(\theta_i)$, $\widehat x^a(\theta_i)$ being the unit 
vector in the direction of  $x^a$ and thus $|\vec x|=C(\mu)^{-1}$.

The adS isometries can be divided into two classes:
quasi-translations and rotations \cite{wein}. We will denote by
${\vec {x}}_{x_0}$ the transform of $\vec {x}$ under a
quasi-translation with parameter $-{\vec {x}}_0$ \cite{wein}: 
\begin{equation}
{\vec {x}}_{x_0}=\vec {x} - {\vec {x}}_0\left[ 
(1+\lambda {\vec {x}}^2)^{1/2} -\frac{[(1+\lambda 
{\vec {x}}_0^2 )^{1/2}-1]}{{\vec {x}}_0^2} (\vec
{x} \cdot {\vec {x}}_0) \right].  
\label{dcuas}
\end{equation}

Note that, when $\lambda=0$, we get ${\vec {x}}_{x_0}=\vec
{x}-{\vec {x}}_0 $, i.e. a translation in flat space. The image
of $\vec {x}$ under an $SO(4)$ rotation $h$ will be called 
\begin{equation}
{\vec {x}}_{h} \equiv R(h) \vec {x},  \label{rot}
\end{equation}
$R(h)$ being a rotation matrix. 

\subsection{Conformal scalar field}

In this subsection, we will calculate the two-point matrix
element for a conformal scalar field, 
\begin{equation}
\langle 0|\phi (x_1)\phi (x_2)|\Psi _\alpha  \rangle,   
\label{scme}
\end{equation}
in the semiclassical approximation. Therefore, we need to find
all the saddle points for the scalar field, $\phi_{x_0h}$, which
are characterised by their position and orientation with respect
to an arbitrary reference. As remarked above, pure adS spacetime
outside a three-sphere can be regarded as the gravitational
saddle point. Then, the conformal scalar field solutions must
satisfy the equation 
\begin{equation}
(\Box +2\lambda )\phi=0, 
 \label{eqsc}
\end{equation}
where $\Box $ is the Laplacian for the adS metric given by Eq.
(\ref{pmetric}) or (\ref{metric}). In order to solve this
equation, we decompose again the scalar field in hyperspherical
harmonics in the three-sphere using polar coordinates, 
\begin{equation}
\phi(x)=a^{-1}(\mu)\sum_{n,\sigma_n}
\chi_{n\sigma _n}(\mu)Q^{n\sigma_n}(\theta_i )
\end{equation}
where $a(\mu)=C(\mu)^{-1}$
is the adS scale factor. In this expression, we have dropped an
irrelevant constant factor $\sqrt{\frac 1{2\pi ^2}}$. For each
mode $n$, we choose a cartesian basis in the corresponding
degenerate space \cite{lifs}, so that
\begin{equation}
Q^{n\sigma _n}(\theta_i)={\cal A}_{a_1\cdots a_{n-1}}
^{n\sigma _n}{\widehat{x}}^{a_1}\cdots {\widehat{x}}^{a_{n-1}},
\end{equation}
where ${\cal A}^{n\sigma _n}$ are the elements of an orthonormal
basis in the linear space of the symmetric tensors of rank $n-1$,
traceless in every pair of indices. Then,
Eq. (\ref{eqsc}) becomes a set of decoupled ordinary differential
equations, one for each coefficient $\chi_{n\sigma _n}$, 
\begin{equation}
{\ddot \chi}_{n\sigma _n}+A(\mu ){\dot \chi}_{n\sigma _n}-
n^2C^2(\mu )\chi_{n\sigma_n}=0.  
\label{eqfn}
\end{equation}
Here, the overdot means derivative with respect to $\mu $ and 
\begin{equation}
A(\mu )=-\dot C(\mu)/C(\mu).
\label{aacc}
\end{equation}
The solutions of these equations must have the same fall-off at
infinity for all $\sigma _n$. This asymptotic behaviour depends
on the choice of vacuum. As second order linear differential
equations, this requirement determines one of the two arbitrary
constants for each $\sigma _n$. The remaining constants are
determined by the value of the radius of the three-sphere that is
going to be identify with a section $\Sigma$ of the wormhole,
$a^0$, and the values of the field, $\chi_{n\sigma _n}^0$, in
this three-sphere. We can write the solutions as
\begin{equation}
\chi_{n\sigma _n}(\mu)=
\frac{\chi_{n\sigma _n}^0}{f_n(\mu^0)}f_n(\mu),
\hspace{10mm}
\mu^0 \equiv \frac{1}{\sqrt{\lambda}}\sinh^{-1}(\sqrt{\lambda}a^0).
\label{mucero}
\end{equation}

The saddle point can then be re-expressed as
\begin{equation}
\phi(x)=C(\mu)\sum_n f_n(\mu){\cal A}^n_{a_1\cdots
a_{n-1}}{\widehat{x}}^{a_1}\cdots {\widehat{x}}^{a_{n-1}},
\label{inter}
\end{equation}
with
\begin{equation}
{\cal A}_{a_1\cdots a_{n-1}}^n=
\sum_{\sigma_n}{\cal A}_{a_1\cdots a_{n-1}}^{n\sigma_n}
\frac{\chi_{n\sigma _n}^0}{f_n(\mu^0)}.
\label{an}
\end{equation}
Note that the tensor ${\cal A}^n$ has the same symmetries as the
elements of the cartesian basis ${\cal A}^{n\sigma _n}$. Equation
(\ref{inter}) is already written in quasi-Euclidean coordinates.
We can obtain all the saddle points for the scalar field by
translating and rotating this particular one: 
\begin{equation}
\phi_{x_0 \bar {h}}(x)=C(\mu)\sum_n f_n(\mu) 
{\cal A}^n_{a_1 \cdots a_{n-1}} 
{\widehat {x}}_{x_0h_n}^{a_1} \cdots {\widehat
{x}}_{x_0h_n}^{a_{n-1}},
\label{egsc}
\end{equation}
where $\bar {h}=\{h_1,\cdots, h_n, \cdots \}$ is the set of
independent rotations for each mode and $\mu=\mu(x,x_0)$ is the
geodesic distance between the points $x$ and $x_0$, defined
through the relation 
\begin{equation}
\mu(x,x_0)=\frac{1}{\sqrt{\lambda}}
\sinh^{-1}(\sqrt{\lambda}|{\vec {x}}_{x_0}|).  
\label{dgeo}
\end{equation}

We are now ready to approximate the matrix element (\ref{scme})
semiclassically:
\begin{eqnarray}
&&\hspace*{-10mm}\langle 0| \phi(x_1) 
\phi (x_2)|\Psi_{N_1,\cdots,N_n,\cdots}\rangle =
\int_0^{+\infty} da \int_{-\infty}^{+\infty}d\chi_1\prod_{m>1} 
\int_0^{+\infty}d\chi_m (\chi_m)^{m^2-1}\times     \nonumber \\
&&\hspace*{-10mm}\int d^4x_0 \sqrt{g(x_0)} \int_{SO(4)} d\bar {h}
\phi_{x_0\bar {h}}(x_1)\phi_{x_0\bar {h}}(x_2) 
\Delta (a^0, \chi^0_n) e^{-I^{sp}(a^0,\chi^0_1)}
\Psi_{N_1,\cdots,N_n,\cdots } (a,\chi_n^0),
\label{meg}
\end{eqnarray}
where $\Delta$ is the sum of the quadratic fluctuations around
the classical solution and the integral over $\bar {h}$
represents and integral over each $h_n$ independently. Note that
the wave function $\Psi$, the action $I^{sp}$ and the determinant
$\Delta$ are rotationally invariant. Thus, the integrand depends
on the group $SO(4)$ only through the saddle point solutions
$\phi_{x_0\bar {h}}$. 

Let us analyse each mode independently. For the lowest mode
$n=1$, the saddle points depend only on $\mu$; therefore, the
average over orientations has no effect and Eq. (\ref{meg})
reduces to
\begin{eqnarray}
\langle 0| \phi(x_1) \phi
(x_2)|\Psi_{N_1}\rangle =&&\hspace*{-6mm}
\int_0^{+\infty} da^0\int_{-\infty}^{+\infty} d\chi^0_1 
\left(\frac{\chi^0_1}{f_1(\mu^0)}\right)^2
\Delta (a^0, \chi^0_1)e^{-I^{sp}(a^0,\chi^0_1)} 
\Psi_{N_1}(a^0, \chi^0_1)\times  \nonumber \\
&&\hspace*{-6mm}\int d^4x_0 \sqrt{g(x_0)} 
C(\mu_1)f_{1}(\mu_1)C(\mu_2)f_{1}(\mu_2).
\label{apscn1}
\end{eqnarray}
The action contains a surface term suitable for fixing an
appropriate variable in the asymptotic region \cite{nos}. We can
choose that variable so that it can be interpreted as an energy.
Different choices of such an energy will lead to the selection of
different vacua for the system, and will require the introduction
of different surface terms in the action. Owing to these terms,
it will not be possible in general to separate the action into
two parts, one for the scale factor and another for the scalar
field. However, there is a particular vacuum for which this
separation is possible: the vacuum in which the associated
solution for the scalar field is the conformal transform of the
only solution damped at infinity when the background is flat; in
this sense, this conformal vacuum is the conformal transform of
the natural vacuum in flat space. Therefore, if $|0\rangle$ in
Eq. (\ref{apscn1}) represents the conformal vacuum, the
dependence on $a^0$ and $\chi_1^0$ in the integral separates. The
part that depends on the conformal scalar field $\chi_1^0$ has
the form
\begin{equation}
\int d\chi_1^0 (\chi_1^0)^2 
{\cal H}_{N_1}e^{-(\chi^0_1)^2}.
\end{equation}
Since the ${\cal H}_{N_1}$'s are the Hermite polynomials, this
integral only gives a non-vanishing contribution when $N_1=0$ or
$N_1=2$. The matrix element for $N_1=0$ represents a
vacuum-to-vacuum transition without any interaction with the
wormhole. Thus, in the conformal vacuum, the wormhole two-point
function for the homogeneous mode $n=1$ finally becomes 
\begin{equation}
\langle 0| \phi(x_1) \phi (x_2)|\Psi_{N_1}\rangle
=\delta_{2,N_1}
 \int d^4x_0 \sqrt{g(x_0)} 
C(\mu_1)f_{1}(\mu_1)C(\mu_2)f_{1}(\mu_2).
\label{frscn1}
\end{equation}
We can interpret $\Psi_{N_1=2}$ as a wormhole state with two
homogeneous scalar particles associated with the conformal
vacuum. A similar analysis applied to three-point and higher
functions would lead us to interpret the basis elements of the
Hilbert space, $\Psi_{N_1}$, as wormhole states containing $N_1$
particles. For any other vacuum choice, there will exist a
different orthonormal basis of the Hilbert space of wormholes
such that only one basis element, apart from the vacuum itself,
will give a non-vanishing contribution to the matrix element
(\ref{frscn1}). In this situation, we could interpret this new
basis element as a wormhole state with two particles associated
with that specific vacuum. Therefore, the interpretation of
wormholes as containing particles has the same ambiguity as the
particle concept in quantum field theory in curved spacetimes. 

Taking into account Eq. (\ref{meg}), the matrix element for the
mode $n=2$ is
\begin{eqnarray}
&&\langle 0| \phi(x_1) \phi (x_2)|\Psi_{N_2}\rangle =
\int_0^{\infty} da^0 \int_0^{\infty}d\chi^0_2 (\chi^0_2)^2 
\Delta (a^0, \chi^0_2) e^{-I^{sp}(a^0, \chi^0_2)} 
\Psi_{N_2}(a^0, \chi^0_2) \times  \nonumber \\
&&\int d^4x_0 \sqrt{g(x_0)} \int_{SO(4)} dh 
C(\mu_1)f_{2}(\mu_1) {\cal A}^{2}_{a_1} 
{\widehat {x}}_{1,x_0h}^{a_1} {\cal A}^{2}_{b_1} 
{\widehat {x}}_{2,x_0h}^{b_1} C(\mu_2)f_{2}(\mu_2). 
\label{apscn2}
\end{eqnarray}
The $SO(4)$ dependence of the integrand comes from the factor 
\begin{equation}
{\cal A}^2_{a_1} {\widehat {x}}_{1,x_0h}^{a_1} 
{\cal A}^2_{b_1} {\widehat {x}}_{2,x_0h}^{b_1},
\end{equation}
because $f_2(\mu_1)$ and $f_2(\mu_2)$ do not depend on the
orientation. This can be re-expressed as 
\begin{equation}
{\cal A}^2_{a_1} R(h)^{a_1}_{\;\; c_1} 
{\widehat {x}}_{1,x_0}^{c_1} {\cal A}^2_{b_1} 
R(h)^{b_1}_{\;\; d_1} {\widehat {x}}_{2,x_0}^{d_1},  
\label{rscn2t}
\end{equation}
where $R(h)^{a_1}_{\;\; c_1}$ denotes a rotation matrix in four
dimensions, i.e. an irreducible unitary representation of the
$SO(4)$ group. Using the property (see, e.g., Ref. \cite{bar}) 
\begin{equation}
\int_{SO(4)} dh R(h)^{a}_{\;\; c}R(h)^{b}_{\;\; d}= 
\delta^{ab}\delta_{cd},
\label{pigr}
\end{equation}
we can straightforwardly carry out the average over orientations,
obtaining 
\begin{equation}
({\cal A}^2 \cdot {\cal A}^2 )
({\widehat {x}}_{1,x_0} \cdot {\widehat {x}}_{2,x_0}).  
\label{rscn2}
\end{equation}
The product ${\cal A}^2_{a_1}{\cal A}^{2\; a_1}$ is proportional
to $(\chi^0_2)^2$, as can be seen from Eq. (\ref{an}). Then, the
matrix element (\ref{apscn2}) gives a non-vanishing result only
in the case $N_2=0$ (vacuum-to-vacuum transition) or $N_2=1$,
that is, when there is a two-particle state associated with the
conformal vacuum inside the wormhole. As in the case of
homogeneous particles, the interpretation of different states of
a wormhole as containing pairs of particles is related to the
vacuum choice. We can write the final result for the matrix
element between the conformal vacuum and a wormhole state as 
\begin{equation}
\langle 0| \phi(x_1) \phi (x_2)|\Psi_{N_2}\rangle =
\delta_{1,N_2}\int 
d^4x_0 \sqrt{g(x_0)} C(\mu_1)f_{2}(\mu_1) 
({\widehat {x}}_{1,x_0} \cdot {\widehat {x}}_{2,x_0})
C(\mu_2)f_{2}(\mu_2) .  
\label{frscn2}
\end{equation}

For an arbitrary mode $n$, the factor that depends on
orientations in Eq. (\ref{meg}) has the form
\begin{equation}
{\cal A}^n_{a_1\cdots a_{n-1}}
{\widehat {x}}_{1,x_0h}^{a_1}\cdots{\widehat {x}}_{1,x_0h}^{a_{n-1}}
{\cal A}^n_{b_1\cdots b_{n-1}}
{\widehat {x}}_{2,x_0h}^{b_1}\cdots{\widehat {x}}_{2,x_0h}^{b_{n-1}}
\label{oav}
\end{equation}
In order to integrate this expression over $SO(4)$, we proceed as
follows. First, we extract the $h$-dependence from each
${\widehat {x}}_{x_0h}^{b}$, i.e.
\begin{equation}
{\widehat {x}}_{x_0h}^{a}=R(h)^{a}_{\;\; b} 
{\widehat {x}}_{x_0}^{b}.
\end{equation}
Then, we absorb this $h$-dependence in the tensor ${\cal A}^n$ by
forming a transformed tensor ${\cal A}^n_h$, which has the same
symmetries as ${\cal A}^n$, because the linear space of tensors
with such symmetries carries an irreducible tensor representation
of the group $SO(4)$ and, therefore, this linear space is stable
under rotations.

Second, using the symmetriser $S_{{\cal A}^n}$ corresponding to
the symmetries of ${\cal A}^n$, we re-express the product ${\cal
A}^n_{h,a_1\cdots a_{n-1}} {\widehat {x}}_{x_0}^{a_1}\cdots
{\widehat {x}}_{x_0}^{a_{n-1}}$ in the form
\begin{equation}
{\cal A}^n_{h,a_1\cdots a_{n-1}}
S_{{\cal A}^n}^{a_1\cdots a_{n-1}},
\label{inter2}
\end{equation}
with $S_{{\cal A}^n}^{a_1\cdots a_{n-1}}= S_{{\cal
A}^n}({\widehat {x}}_{x_0}^{a_1} \cdots {\widehat
{x}}_{x_0}^{a_{n-1}})$. As an illustration, for the mode $n=3$,
$S_{{\cal A}^n}^{a_1a_2}$ is given by
\begin{equation}
S_{{\cal A}^n}^{a_1a_2}= \left({
\widehat {x}}_{x_0}^{a_1}{\widehat {x}}_{x_0}^{a_2} -\frac{1}{4}\delta^{a_1
a_2}\right).  
\label{rscn3}
\end{equation}
Expression (\ref{inter2}) is a scalar product in the linear space
of tensors with the symmetries of ${\cal A}^n$. We will write it
in the form ${\cal A}^n_h \cdot S_{{\cal A}^n}$. Since ${\cal
A}^n_h= R(h){\cal A}^n$, with $R(h)$ being in the appropriate
irreducible representation of $SO(4)$, and 
\begin{equation}
\int_{SO(4)}dh R(h) \otimes R(h)= 1 \otimes 1,
\end{equation}
where $\otimes$ denotes the tensor product, the average over
orientations of Eq. (\ref{oav}) results in
\begin{equation}
({\cal A}^n \cdot {\cal A}^n)
\left(S_{1,{\cal A}^n} \cdot S_{2,{\cal A}^n}\right).
\end{equation}
The term ${\cal A}^n \cdot {\cal A}^n$ is again proportional to
$(\chi_n^0)^2$. Then, the matrix element for any mode $n>1$ in
the conformal vacuum is
\begin{equation}
\langle 0| \phi(x_1) \phi (x_2)|\Psi_{N_n}\rangle =
\delta_{1,N_n} \int d^4x_0 \sqrt{g(x_0)} C(\mu_1)f_{n}(\mu_1) 
\left(S_{1,{\cal A}^n} \cdot S_{2,{\cal A}^n}\right)
C(\mu_2)f_{n}(\mu_2).  
\label{frscg}
\end{equation}

Finally, the matrix element (\ref{meg}) can be written as the sum
of the calculated contributions of each mode. Indeed, the
additional terms that involve cross products of different modes
do not give any contribution because, in these cases, the average
over orientations contains the vanishing integral 
\begin{equation}
\int_{SO(4)}dh R(h)\otimes 1=0.
\end{equation}

\subsection{Electromagnetic field}

The matrix element for the electromagnetic wormhole interaction
is given by the gauge invariant quantity 
\begin{equation}
\langle 0|A_{\mu}(x_1) A_{\nu}(x_2)|\Psi_\alpha\rangle.
\label{elme}
\end{equation}
We choose the gauge in which the only degrees of freedom are the
transverse waves and deal with each mode
$A_i^n=\sum_{\sigma_n}\beta_{n\sigma_n} S_i^{n\sigma_n}$
separately. The Fadeev-Popov determinant associated with this
gauge fixing condition is independent of the fields and can be
reabsorbed in the normalisation \cite{fadpov}. The saddle point
for the electromagnetic potential in adS spacetime must satisfy
the equation 
\begin{equation}
( \Box +3\lambda ) A^n_{\mu}=0.
\label{eqel}
\end{equation}
In order to find solutions to this equation, we will follow the
same steps as in the case of the scalar field. We first choose a
cartesian basis in the degenerate space corresponding to each
mode \cite{lifs}: 
\begin{equation}
A_i^n=\sum_{\sigma_n}\beta_{n\sigma_n}(\mu) \frac{\partial
{\widehat {x}}^{a_1}}{\partial x^i} {\cal B}^{ n\sigma_n}_{a_1
a_2;a_3 \cdots a_{n}} {\widehat {x}}^{a_2} {\widehat
{x}}^{a_3}\cdots {\widehat {x}}^{a_{n}}.  \label{A1}
\end{equation}
Here, the $x_i$'s are coordinates in the three-sphere and 
the functions $\beta_{n\sigma_n}$ satisfy the equation 
\begin{equation}
{\ddot {\beta}}_{n\sigma_n} + A{\dot {\beta}}_{n\sigma_n} - n^2
C^2 \beta_{n\sigma_n} =0,  \label{eqbn}
\end{equation}
with $A(\mu)$ being given in Eq. (\ref{aacc});
${\cal B}^{n\sigma_n}_{a_1a_2;a_3 \cdots a_{n}}$ are the elements
of an orthonormal basis in the space of tensors antisymmetric in
their first two indices, symmetric with respect to all other
indices, and that vanish when contracting any pair of indices or
when taking a cyclic sum over $a_1, a_2$ and any other index. As
we did with the scalar field, we will isolate all the
$\mu$-dependence in just one variable for each mode,
\begin{equation}
\beta_{n\sigma_n}(\mu)=\frac{\beta^0_{n\sigma_n}}{f_n(\mu^0)}
f_n(\mu),
\end{equation}
where $\mu^0$ was defined in Eq. (\ref{mucero}). We then define a
new tensor, 
\begin{equation}
{\cal B}^n_{a_1 a_2;a_3 \cdots a_{n}}=\sum_{\sigma_n}
\frac{\beta^0_{n\sigma_n}}{f_n(\mu^0)} 
{\cal B}^{n\sigma_n}_{a_1 a_2;a_3 \cdots a_{n}},
\end{equation}
which contains all the dependence on the arguments of the
electromagnetic wormhole wave function.

The vector $A_i^n$ on the three-sphere can be expressed as a
vector on adS spacetime by contracting it  with 
$\frac{\partial x_i}{\partial x^{\mu}}$: 
\begin{equation}
A^n_{\mu}(x)=f_{n}(\mu)\frac{1}{|\vec {x}|}
{\cal B}^{n}_{a_1a_2;a_3 \cdots a_n} 
(\delta_{\mu}^{a_1}-{\widehat {x}}^{a_1}
{\widehat {x}}^b\delta_{\mu b})
{\widehat {x}}^{a_2}{\widehat {x}}^{a_3}\cdots 
{\widehat {x}}^{a_n}.
\end{equation}
Owing to the antisymmetry of the tensor ${\cal B}^{n}$, we can
sustitute the term inside the bracket by
\begin{equation}
e_{\mu}^{a_1}\equiv
\delta_{\mu}^{a_1}-
\frac{[(1+\lambda{\vec {x}}^2)^{1/2}-1]}
{(1+\lambda{\vec {x}}^2)}
{\widehat{x}}^{a_1}{\widehat{x}}^{a}\delta_{a\mu}.
\end{equation}
In this way, the tensor character of ${\cal B}^{n}_{a_1a_2;a_3
\cdots a_n}$ under local changes of coordinates shows up
explicitly. The tetrad $e_{\mu}^{a}$ defined here satifies the
following relations, which we will use in what follows:
\begin{equation}
\frac{\partial x_{h}^{\nu}}{\partial x^{\mu}} e_{\nu}^{a} 
({\vec {x}}_{h})=R(h)^{a}_{\;\; b}e_{\mu}^{b},
\label{prot}
\end{equation}
\begin{equation}
\frac{\partial x_{x_0}^{\nu}}{\partial x^{\mu}} e_{\nu}^{a} 
({\vec {x}}_{x_0})= e_{\mu}^{ a},  
\label{ptrans}
\end{equation}
\begin{equation}
g_{\mu \nu}= e_{\mu}^{ a} e_{\nu}^{ b} \delta_{ab}.
\label{pinloc} 
\end{equation}
The saddle point can still be separated into positive and
negative helicity parts. This can be achived by introducing 
\begin{equation}
{\cal B}^{n\pm}=\frac{1}{2}\left({\cal B}^{n} 
\pm {^{*}{\cal B}}^{n}\right),
\end{equation}
where $ {^{*}{\cal B}}^{n}$ is the dual of the tensor ${\cal
B}^{n}$ formed by contracting it with the Levi-Civita tensor
\begin{equation}
{^{*}{\cal B}}^{n}_{a_1a_2;a_3 \cdots a_n}=
\epsilon_{a_1a_2}^{\;\;\;\;\;\;\;cd}
{\cal B}^{n}_{cd;a_3 \cdots a_n}.
\end{equation}

The transformed saddle point for the mode $n$ and positive or
negative helicity can be written, after some manipulations
involving properties (\ref{prot}) and (\ref{ptrans}) of the
tetrad, as
\begin{equation}
(A_{x_0h}^{n\pm})_{\mu}(x)=f_{n}(\mu) \frac{1}{|{\vec
{x}}_{x_0}|} {\cal B}^{n\pm}_{h,a_1a_2;a_3 \cdots a_n}
e_{\mu}^{a_1} {\widehat {x}}_{x_0}^{a_2} {\widehat
{x}}_{x_0}^{a_3}\cdots {\widehat {x}}_{x_0}^{a_n},
\label{elsp}
\end{equation}
where we have absorbed all the $h$ dependence in the transformed
tensor ${\cal B}^{n\pm}_{h}$. Once we notice that the ${\cal
B}^{n\pm}$'s carry an irreducible tensor representation of
$SO(4)$ (as the ${\cal A}^n$'s did in the scalar case), we can
apply the same steps discussed in the previous subsection to
carry out the average over orientation. The result of this
average is proportional to
\begin{equation}
({\cal B}^{n\pm}\cdot {\cal B}^{n\pm})
(S_{1,{\cal B}^{n\pm} \mu} \cdot 
S_{2,{\cal B}^{n\pm} \nu}),
\end{equation}
with $S_{{\cal B}^{n\pm} \mu}^{a_1a_2;a_3\cdots a_n}= 
S_{{\cal B}^{n\pm}}(e_{\mu}^{ a_1}{\widehat {x}}_{x_0}^{a_2}
{\widehat {x}}_{x_0}^{a_3}\cdots {\widehat {x}}_{x_0}^{a_n})$
being the symmetriser associated with the symmetries of
${\cal B}^{n\pm}$. 

As in the inhomogeneous scalar field case, the matrix element
between the conformal vacuum and a wormhole state is non-zero
only when $N_n^{\pm}=1$, i.e. when the wormhole contains a
two-photon state in the mode $n$ with positive or negative
helicity associated with the conformal vacuum, since the term
${\cal B}^{n\pm}\cdot {\cal B}^{n\pm}$ is proportional to
$(\beta_n^{\pm0})^2$.

At last, we can write the electromagnetic wormhole matrix element
as 
\begin{equation}
\langle 0|A_{\mu}(x_1) A_{\nu}(x_2)|\Psi_{N_n^{\pm}}\rangle
=\delta_{1,N_n^{\pm}}
 \int d^4x_0 \sqrt{g(x_0)}  C(\mu_1)f_n(\mu_1)
(S_{1,{\cal B}^{n\pm}\mu}\cdot S_{2,{\cal B}^{n\pm} \nu})
C(\mu_2)f_n(\mu_2).
\label{frel}
\end{equation}
For the lowest mode $n=2$, the term inside the bracket is the
Euclidean contraction of two tensors $S_{{\cal B}^{2\pm}
\mu}^{a_1a_2}$ of the form
\begin{equation}
S_{{\cal B}^{2\pm} \mu}^{a_1a_2}=\frac{1}{2}\left[
(e_{\mu}^{ a_1}{\widehat {x}}_{x_0}^{a_2}-
e_{\mu}^{ a_2}{\widehat {x}}_{x_0}^{a_1}) \pm
\epsilon^{a_1a_2}_{\;\;\;\;\;\;\;cd}
(e_{\mu}^{ c}{\widehat {x}}_{x_0}^{d}-
e_{\mu}^{ d}{\widehat {x}}_{x_0}^{c})\right],
\end{equation}
one evaluated at the point $x_1$ and the other at the point
$x_2$.

These results have been obtained for the conformal vacuum, to
which the wormhole basis (\ref{emwf}) is related, although it is
clear that we could have performed an equivalent analysis for any
other choice of vacuum.

\subsection{Gravitons}

The techniques explained in the previous subsections can also be
applied to gravitons in adS spacetime. In this case, the
two-point funtions are 
\begin{equation}
\langle 0|h_{\mu \nu}(x_1) h_{\rho \sigma}(x_2)|\Psi_\alpha\rangle.
\label{grme}
\end{equation}
The only non-vanishing elements of the metric perturbations, upon
a suitable gauge fixation, are
$h_{ij}=\sum_{n,\sigma_n}d_{n\sigma_n}G^{n\sigma_n}_{ij}$. The
equation for $h_{\mu \nu}$ reads
\begin{equation}
(\Box +2\lambda)h_{\mu \nu}=0.  \label{eqgr}
\end{equation}
Using a cartesian basis in the degenerate space corresponding
to the mode $n$ \cite{lifs}, we can write
\begin{equation}
h_{ij}=f_n(\mu) \frac{\partial {\widehat {x}}^{a_1}}{\partial
x^i} \frac{ \partial {\widehat {x}}^{a_3}}{\partial x^j} {\cal
C}^{ n}_{a_1 a_2;a_3 a_4; a_5 \cdots a_{n+1}} {\widehat
{x}}^{a_2} {\widehat {x}}^{a_4} {\widehat {x}}^{a_5} \cdots
{\widehat {x}}^{a_{n+1}},
\end{equation}
where we have already isolated the $\mu$ dependence in just one
variable $f_n$ satisfying now the equation
\begin{equation}
{\ddot {f}}_n - A{\dot {f}}_n -
(6A^2-4AC+C^2(n^2-3))f_n=0.  \label{eqdn}
\end{equation}

The ${\cal C}^{ n}$'s, which contain the dependence on the
arguments of the wave function, are tensors with the following
properties: they are antisymmetric with respect to each pair of
indices $a_1 a_2$ and $a_3 a_4$; symmetric with respect to all
other indices, as well as under the interchange of the pair $a_1
a_2$ with the pair $a_3 a_4$; finally, they vanish when
contracting any two indices or under cyclic sums over triplets of
indices that contain the pair $a_1 a_2$ or $a_3 a_4$. 

Now, we can find the transformed saddle point for the mode $n$
and positive or negative helicity:
\begin{equation}
(h_{x_0h}^{n\pm})_{\mu \nu}(x)=f_{n}(\mu) \frac{1}{|{\vec
{x}}_{x_0} |^2} {\cal C}^{n\pm}_{h,a_1 a_2;a_3 a_4;a_5 \cdots
a_{n+1}} e_{\nu}^{ a_1} {\widehat {x}}_{x_0}^{a_2} e_{\mu}^{ a_3}
{\widehat {x}}_{x_0}^{a_4} {\widehat {x}}_{x_0}^{a_5} \cdots
{\widehat {x}}_{x_0}^{a_{n+1}},
\end{equation}
where 
\begin{equation}
{\cal C}^{n\pm}=({\cal C}^{n} \pm {^{*}{\cal C}}^{n}),
\end{equation}
\begin{equation}
{^{*}{\cal C}}^{n}_{a_1 a_2;a_3 a_4;a_5 \cdots a_{n+1}}=
\frac{1}{2}(\epsilon_{a_1a_2}^{\;\;\;\;\;\;\;cd}
{\cal C}^{n}_{cd;a_3 a_4;a_5 \cdots a_{n+1}}+
\epsilon_{a_3a_4}^{\;\;\;\;\;\;\;cd}
{\cal C}^{n}_{cd;a_1 a_2;a_5 \cdots a_{n+1}}).
\label{gduals}
\end{equation}
Similarly to what happened in the electromagnetic case, the ${\cal
C}^{n\pm}$'s carry an irreducible tensor representation of the
rotation group and, therefore, the average over orientations
gives the result
\begin{equation}
({\cal C}^{n\pm} \cdot {\cal C}^{n\pm})
(S_{1,{\cal C}^{n\pm} \mu \nu} \cdot S_{2,{\cal C}^{n\pm} \rho
\sigma}),
\end{equation}
in which we have introduced the symmetriser
\begin{equation}
S_{{\cal C}^{n\pm} \mu \nu}^{a_1 a_2;a_3 a_4;a_5 \cdots a_{n+1}}=
S_{{\cal C}^{n\pm}}(e_{\mu}^{ a_1}{\widehat
{x}}_{x_0}^{a_2} e_{\nu}^{ a_3}{\widehat
{x}}_{x_0}^{a_4}{\widehat {x}}_{x_0}^{a_5}
\cdots {\widehat {x}}_{x_0}^{a_{n+1}}).
\end{equation}

The matrix elements between a vacuum and the elements
$\Psi_{\alpha}$ of a basis of graviton wormhole states are
\begin{eqnarray}
&&\langle 0|h_{\mu \nu}(x_1) h_{\rho
\sigma}(x_2)|\Psi_{\alpha}\rangle =\nonumber\\
&&K(0,\Psi_{\alpha}) \int d^4x_0 \sqrt{g(x_0)} 
C^2(\mu_1)f_n(\mu_1)
(S_{1,{\cal C}^{n\pm} \mu \nu} \cdot S_{2,{\cal C}^{n\pm} \rho
\sigma})C^2(\mu_2)f_n(\mu_2),
\end{eqnarray}
where $K(0,\Psi_{\alpha})$ is a constant that depends on the
vacuum $|0\rangle$ and the state $\Psi_{\alpha}$. The vacuum
$|0\rangle$ is specified by its associated mode expansion
\cite{wada}. It is worth noting that, in general, these modes do
not fall off at infinity \cite{tur} and, thus, the back-reaction
effects of gravitons on the background diverge. Since there
exists graviton creation in this system, we cannot interpret the
states of the Hilbert space as wormholes containing certain
number of gravitons.

Before we end this section, let us write the symmetriser 
$S_{{\cal C}^{n}}$ for the lowest mode $n=3$ as an example:
\begin{equation}
S_{{\cal C}^n \mu \nu}^{a_1a_2a_3a_4} =R_{\mu \nu}^{a_1 a_2 a_3
a_4} + \left(\delta^{a_1[a_4}R_{\mu \nu}^{a_3 ]a_2}+
\delta^{a_2[a_3}R_{\mu \nu}^{a_4]a_1 }\right) +\frac{1}{3}R_{\mu
\nu} \left(\delta^{a_1 [a_3}\delta^{a_4] a_2}\right),
\label{rgrn3}
\end{equation}
with 
\begin{eqnarray}
R_{\mu \nu}^{a_1 a_2 a_3 a_4}&=&{\widehat {x}}_{x_0}^{a_1}
{\widehat {x}} _{x_0}^{a_3} \left(e_{\mu}^{ a_2}e_{\nu}^{ a_4} +
e_{\mu}^{ a_4} e_{\nu}^{ a_2} \right)- {\widehat {x}}_{x_0}^{a_1}
{\widehat {x}} _{x_0}^{a_4} \left(e_{\mu}^{ a_3}e_{\nu}^{ a_2}
+e_{\mu}^{ a_2}e_{\nu}^{ a_3} \right)- \nonumber \\
&&{\widehat {x}}_{x_0}^{a_2} {\widehat {x}}_{x_0}^{a_3}
\left(e_{\mu}^{ a_1} e_{\nu}^{ a_4}+ e_{\mu}^{ a_4} e_{\nu}^{
a_1} \right)+ {\widehat {x}}_{x_0}^{a_2} {\widehat
{x}}_{x_0}^{a_4} \left(e_{\mu}^{ a_1} e_{\nu}^{ a_3}+ e_{\mu}^{
a_3} e_{\nu}^{ a_1}\right), \nonumber \\[2mm]
 R_{\mu \nu}^{a_2 a_4}&=&
e_{\mu}^{ a_2}e_{\nu}^{ a_4}+ e_{\mu}^{ a_4}e_{\nu}^{
a_2}+2g_{\mu \nu} {\widehat {x}}_{x_0}^{a_2} {\widehat {x}}
_{x_0}^{a_4}- \\
&&n_{\mu}e_{\nu}^{ a_2} {\widehat {x}}_{x_0}^{a_4}
-n_{\nu}e_{\mu}^{ a_2} {\widehat {x}}_{x_0}^{a_4}
-n_{\mu}e_{\nu}^{ a_4} {\widehat {x}} _{x_0}^{a_2}
-n_{\nu}e_{\mu}^{ a_4} {\widehat {x}}_{x_0}^{a_2}, \nonumber
\\[2mm]
R_{\mu \nu}&=&4(g_{\mu \nu}-n_{\mu}n_{\nu}).\nonumber
\end{eqnarray}
In this expression, $n_{\mu} \equiv e_{\mu a}{\widehat
{x}}_{x_0}^{a}$ is the normal in $\vec {x}$ to the three-sphere
centered in ${\vec {x}}_0$.

\section{ Effective interactions}

\setcounter{equation}{0}
\label{ei}

It has been proposed \cite{col,haw1} that the effects of baby
universes branching off and joining onto a large universe can be
modelled by adding local effective interaction terms to a quantum
field theory in a fixed background. In this section, we
will search for interaction Lagrangians ${\cal L}_I^{\alpha}$
that, via the formula 
\begin{equation}
\langle 0|\Phi(x_1)\Phi(x_2) \int d^4x_0 \sqrt{g(x_0)} {\cal L}
_I^{\alpha} \left(\Phi (x_0)\right)|0\rangle,  
\label{ime}
\end{equation} 
provide results equivalent to those calculated in Sec. \ref{tpf}.
In this formula, $|0\rangle$ represents a matter field vacuum in
adS background.

We will use the following ansatz for the interaction Lagrangian: 
\begin{equation}
{\cal L}_I= \Theta\Phi \cdot \Theta\Phi,
\end{equation}
where $\Theta$ is a linear operator constructed with covariant
derivatives $\nabla_{\rho}$, the adS metric $g_{\rho\sigma}$ and
the Levi-Civita tensor. The dot in this expression means
contraction of indices with the metric $g_{\rho\sigma}$. If we
contract each spacetime index in $\Theta\Phi$ with the tetrads
$e_{\rho }^{ a}$, we transform a tensor under general coordinate
transformations into a tensor under local changes of coordinates
and the dot becomes Euclidean contraction.

Then, the linearity of the operator $\Theta$ and Wick's theorem
allow us to write Eq. (\ref{ime}) in the form
\begin{equation}
\int d^4x_0 \sqrt{g(x_0)}
\Theta'G(x_1,x_0) \cdot \Theta'G(x_2,x_0),
\label{effint}
\end{equation}
where $G(x,x_0)$ is the propagator of the matter field $\Phi$.
This propagator and all its covariant derivatives are bitensors
invariant under the isometries of adS spacetime \cite{allen1}.
Primed indices (as well as the prime in $\Theta'$) will be
associated with the tangent space at the point $x_0$ of the adS
background, while unprimed indices will be associated with the
tangent space at the point $x$. The parallel transporter
$g_{\mu\rho'} \equiv e_{\mu}^{ a}e_{\rho'}^{ a'} \delta_{aa'}$
connects these two tangent spaces. In this section, we will
compare expression (\ref{effint}) with the matrix elements
obtained in the previous section for a wormhole interacting with
a conformal scalar field, photons and gravitons. From this
comparison, we will deduce the form of the linear operator
$\Theta$ and, consequently, the effective interaction Lagrangian
for each matter content.

\subsection{Conformal scalar field}

The propagator for a conformal scalar field in adS spacetime is an
invariant biscalar satisfying the equation 
\begin{equation}
(\Box +2\lambda )G(x,x_0)=0.
\end{equation}
It depends only on the geodesic distance $\mu (x,x_0)$ between
the two points $x$ and $x_0$, an invariant quantity under the
whole group of isometries of the adS geometry \cite{allen1}. The
explicit form of this propagator is determined by the choice of
vacuum.

Notice that $G(\mu)$ is proportional to the homogeneous saddle
point of the field $\phi$, since both of them depend only on
$\mu$ and satisfy the same equation, namely, $\ddot {G}+3A\dot
{G}+ 2\lambda G=0$. We can then write the two-point function for
the homogeneous modem, Eq. (\ref{frscn1}), as
\begin{equation}
\int d^4x_0 \sqrt{g(x_0)} G(\mu_1)G(\mu_2).
\end{equation}
Comparison with expression (\ref{effint}) shows that, in this
case, $\Theta$ is the identity operator. Thus, the effective
interaction Lagrangian that reproduces the effects of wormholes
with two particles in the homogeneous mode is ${\cal
L}_I(\phi)=\phi^2$.

Let us now analyse the mode $n=2$. It is convenient to introduce
the normal at the point $x_0$ to the three-sphere centered in
$x$, $n_{\rho'} \equiv -e_{\rho'a'} {\widehat {x}}_{x_0}^{a'}$,
analogous to the normal $n_{\mu} \equiv e_{\mu a} {\widehat
{x}}_{x_0}^{a}$ defined in Sec. \ref{tpf}. Naturally, these two
normals are the covariant derivatives of the geodesic distance
with respect $x_0$ and $x$, that is, $n_{\rho^{\prime}}=
\nabla_{\rho^{\prime}}\mu$ and $n_{\mu}=\nabla_{\mu}\mu$. In view
of these comments, it is worth noting that $-{\widehat
{x}}_{x_0}^{a'}$ and ${\widehat {x}}_{x_0}^{a}$ are the normals
$n^{a'}$ and $n^a$ in local coordinates. The parallel transporter
in local coordinates is just $\delta_{aa'}$, so the primed or
unprimed character of the local coordinates is irrelevant. 

It is easy to see that 
\begin{equation}
e_{\rho'}^{ a'}\nabla^{\rho'}G(\mu)=- \dot {G}(\mu){\widehat
{x}}_{x_0}^{a'}
\end{equation}
and, moreover, that $\dot {G}$ satisfies the same equation as 
$C(\mu)f_2(\mu)$.
Then, a term of the form
\begin{equation}
\int d^4x_0 \sqrt{g(x_0)} 
\nabla'G(x_1,x_0) \cdot \nabla'G(x_2,x_0) 
\end{equation}
reproduces the result obtained for the matrix element
(\ref{frscn2}) with a single two-particle state in the mode $n=2$
inside the wormhole. Comparison with Eq. (\ref{effint}) shows
that $\Theta\phi=\nabla\phi$ and, thus, the effective interaction
Lagrangian is $\nabla^{\rho}\phi\nabla_{\rho}\phi$.

For an arbitrary mode $n$, we need to find an operator $\Theta_n$
that, acting on the propagator, reproduces each of the two
factors of Eq. (\ref{frscg}), i.e. $\Theta_n'G=C(\mu)f_n(\mu)
S_{{\cal A}^n}$. Such operator turns out to be the only symmetric
traceless linear combination with constant coefficients of all
possible terms with $n-1$ indices that we can form using the
primed metric and $n-1$ or fewer primed covariant derivatives.
Let us prove this statement. The requirement that $\Theta_n'$ be
symmetric and traceless ensures that $\Theta_n'G$ is the product
of $S_{{\cal A}^n}$ and some function that depends only on the
geodesic distance, denoted by $C(\mu)F_n(\mu)$ for convenience.
This is so because $\Theta_n'G$ is an invariant bitensor and,
thus, it must be constructed only with the normal
$n^{a'}=-{\widehat {x}}_{x_0}^{a'}$ and the metric
$\delta^{a'b'}$ \cite{allen1} in the same way as $S_{{\cal A}^n}$
was. We can also explicitly check that $\Theta_n'G$ can be
expressed in the form $C(\mu)F_n(\mu)S_{{\cal A}^n}$ by employing
the relation between the derivative of the normal, the normal
itself and the metric, which in local coordinates reads
\begin{equation}
\nabla_{a'}n_{b'}=A(\mu)(\delta_{a'b'}-n_{a'}n_{b'}).
\end{equation}
We will now show by induction that $F_n(\mu)$ satisfies the 
same equation as $f_n(\mu)$, 
\begin{equation}
\ddot f_n+A\dot f_n-n^2 C^2 f_n=0,
\end{equation}
for any mode $n$. We have already seen that this is true for
$n=1$. Let us assume that $\Theta_n'G=C(\mu)f_n(\mu)S_{{\cal
A}^n}$. Then, $\Theta_{n+1}'G$, given by the symmetric traceless
covariant derivative of $\Theta_{n}'G$, becomes after some
manipulations $C^{1-n}\frac{d}{d\mu}(C^n f_n) S_{{\cal
A}^{n+1}}$. Finally, it is not difficult to check that
$F_{n+1}=C^{-n} \frac{d}{d\mu}(C^n f_n)$ satisfies the same
equation as $f_{n+1}$.

The application of this general prescription to the mode $n=3$
provides the operator $\Theta_3^{ab}=\nabla^{a}\nabla^{b}
+\frac{1}{2}\lambda \delta^{ab}$. The interaction term associated
with a wormhole containing a pair of scalar particles in the mode
$n=3$ is, therefore,
\begin{equation}
\left(\nabla^{\rho}\nabla^{\sigma}\phi
+\frac{1}{2}\lambda
g^{\rho\sigma}\phi\right)
\left(\nabla_{\rho}\nabla_{\sigma}\phi
+\frac{1}{2}\lambda
g_{\rho\sigma}\phi\right).
\end{equation}

\subsection{Electromagnetic field}

The equation for the electromagnetic propagator in an adS
background reads 
\begin{equation}
\left[g_{\mu\nu}( \Box +3\lambda ) +
(\zeta-1)\nabla_{\mu}\nabla_{\nu}\right] G^{\nu\rho^{\prime}}=0,
\end{equation}
where $\zeta$ is a constant that specifies the gauge choice. For
an arbitrary mode $n$, the bitensor
$(\Theta_n'G)_{\mu}^{a_1'a_2';a_3' \cdots a_n'}$ is given by a
linear combination of terms of the form
\begin{equation}
g^{a_n'a_{n-1}'} \cdots g^{a_{l+1}'a_{l}'}\nabla^{a_{l-1}'}
\cdots\nabla^{a_2'}G_{\mu}^{a_1'}.
\end{equation}
The primed index structure of $\Theta_n'G$ must be the same as
that of $S_{{\cal B}^n \mu}^{a_1'a_2';a_3' \cdots a_n'}$, i.e. it
must be antisymmetric in its first two indices, symmetric with
respect to all other indices and must vanish when contracting any
pair of indices or when taking a cyclic sum over $a_1$, $a_2$ and
any other index. The proof of this statement goes along the same
lines as that for the scalar field. The symmetry of $\Theta_n'G$
ensures that $\Theta_n'G=C(\mu)F_n(\mu)S_{{\cal B}^n}$ for some
function $F_n(\mu)$. This funtion $F_n$ satisfies the same
equation as the electromagnetic saddle point function $f_n$,
provided that $F_{n-1}$ does, because $\Theta_n'$ is constructed
from $\Theta_{n-1}'$ by taking its symmetric traceless covariant
derivative as in the the scalar field case. It only remains to
check that this is true for the lowest mode $n=2$. In this case,
\begin{equation}
(\Theta_2'G)_{\mu}^{a'b'}=\nabla^{a'}G_{\mu}^{b'}-
\nabla^{b'}G_{\mu}^{a'}
\label{fmu}
\end{equation}
and can be rewritten as
\begin{equation}
(\Theta_2'G)_{\mu}^{a'b'}=C(\mu)F_2(\mu)
(e_{\mu}^{ a'}{\widehat {x}}_{x_0}^{b'}-
e_{\mu}^{ b'}{\widehat {x}}_{x_0}^{a'})
\end{equation}
for some function $F_2(\mu)$. We define a new object
$(\Theta_2'G)_{\mu\nu}^{a'b'}=
\nabla_{\nu}(\Theta_2'G)_{\mu}^{a'b'}-
\nabla_{\mu}(\Theta_2'G)_{\nu}^{a'b'}$, which is equal to
$\langle 0|F_{\mu\nu}(x) F^{a'b'}(x_0)|0\rangle$, since
$G_{\mu\rho'}=\langle 0|A_{\mu}(x) A^{\rho'}(x_0)|0\rangle$.
Therefore, it is gauge invariant and satisfies the field equation
$\nabla^{\nu}(\Theta_2'G)_{\mu\nu}^{a'b'}=0$. This translates
into a second order differential equation for $F_2$, which is
precisely the equation satisfied by $f_2(\mu)$ \cite{allen1}.
 
The interaction Lagrangian for the mode $n=2$ and positive and
negative helicity can be obtained from the self-dual and
anti-self-dual parts, $\Theta_{2+}$ and $\Theta_{2-}$, of the
operator $\Theta_{2}$ given in Eq.. (\ref{fmu}). These Lagrangians
are
\begin{equation}
(\Theta_{2\pm}A)^2=(F \pm{^{*}F})^2
\end{equation}
Since $F_{\rho\sigma}{^*F}^{\rho\sigma}$ is a topological
invariant, both interaction Lagrangians reduce to $F^2$.
Similarly, for an arbitrary mode $n$, the effective interaction
terms corresponding to both helicities, $(\Theta_{n\pm}A)^2$,
coincide because $\Theta_n A \cdot {^*\Theta_n} A$ can be seen to
be a topological invariant.

\subsection{Gravitons}

\setcounter{equation}{0}

In this case, $\Theta_n'G$ is a bitensor constructed with metrics
and covariant derivatives with the same index structure and
symmetries as the $S_{{\cal C}^n}$'s. Owing to this structure, it
has the form $C(\mu)^2F_n(\mu)S_{{\cal C}^n}$ for some function
$F_n(\mu)$. In order to prove that $F_n(\mu)$ satisfies Eq.
(\ref{eqdn}), we first note that $\Theta_n'G=\langle 0|h(x)
\Theta_n' h(x_0)|0\rangle$. In the gauge chosen in Sec. 3,
$\Theta_n'G$ satisfies
\begin{equation}
(\Box +2\lambda)\Theta_n'G=0,
\label{box}
\end{equation}
because $ (\Box +2\lambda) h_{\mu\nu}=0$. Noting that $S_{{\cal
C}^n \mu\nu}^{a_1'a_2';a_3'a_4';a_5' \cdots a_n'}$ are just the
transverse traceless tensor harmonics $G_{ij}^{n\sigma_n}$
expressed in a cartesian basis for the degenerate space of mode
$n$ and in spacetime coordinates, Eq. (\ref{box}) becomes an
equation for $F_n$, which is obviously Eq. (\ref{eqdn}). The
interaction Lagrangian is then given by $(\Theta_{n\pm}h)^2$ for
positive and negative helicities.

For the lowest mode $n=3$, $ \Theta_{3}h$ is the Weyl tensor of
the perturbation $h_{\mu\nu}$ (the Weyl tensor of the adS
background vanishes). The Lagrangians associated with both
helicities reduce then to $C_{\rho\sigma\eta\delta}
C^{\rho\sigma\eta\delta}$, because the cross term
$C_{\rho\sigma\eta\delta} {^*}C^{\rho\sigma\eta\delta}$ is equal
to the Pontryagin invariant \cite{lafl}.

\section{Summary and conclusions}

\setcounter{equation}{0}
\label{sc}

In this work, we have studied the effects of wormholes in
low-energy physics in an adS background. We have calculated the
wormhole matrix element between a vacuum and an arbitrary
wormhole state, $\langle 0|\Phi(x_1)\Phi(x_2)|\Psi_\alpha\rangle$, 
explicitly and have found an effective interaction Lagrangian that
reproduces this matrix element within the context of quantum
field theory in adS background. As a first step, it has been
necessary to find the Hilbert space of wormhole states. The
invariance under the isotropy group $SO(4)$ only permits the
existence of wormholes containing rotationally invariant matter
states. For the inhomogeneous modes, this means that
single-particle states are not allowed. We have considered scalar
and electromagnetic fields, which are conformally coupled to
gravity, and gravitons, which are not, because of the presence of
a cosmological constant. In this case, the WDW equation cannot be
separated but the wormhole Hilbert space can still be formally
constructed in the semiclassical approximation.

We have seen
that, for each vacuum choice, there exists an orthonormal basis
in the Hilbert space of wormholes such that this matrix element
vanishes for all basis states except for one of them and the
vacuum itself. A similar analysis applies to three-point and
higher functions. So, wormhole basis states can be labelled by
the number of particles that they contain. Thus, the ambiguity in
the choice of vacuum which is present in quantum field theory in
curved spacetimes also shows up in wormhole physics. In the
graviton case, we cannot label the wormhole states by their
particle content because there is creation of gravitons owing to
the expansion of the asymptotic region. The states are global
states background-gravitons.
 
For each inhomogeneous mode $n$, the wormhole-induced 
effective interaction
Lagrangian in adS spacetime is of the form
\begin{equation}
{\cal L}_I=(\Theta_n\Phi)^2,
\end{equation}
where the linear operator $\Theta_n$ can be
constructed recursively by symmetrising $\nabla\Theta_{n-1}$ and
substracting all its traces. 
Furthermore, for photons
and gravitons, we have to deal with both helicities, positive and
negative, separately. This amounts to introducing operators
$\Theta_{n\pm}$ which are the self-dual and anti-self-dual parts
of $\Theta_{n}$. We obtain the same interaction Lagrangian
for positive and negative helicities, as expected, because the
cross product $\Theta_n\Phi \cdot {^{*}\Theta}_n\Phi$ is a
topological invariant, as can be checked by direct calculation.
The operator $\Theta_{n_0}$ for the
lowest mode is specific to each matter content: it is the
identity for the scalar field, the antisymmetric derivative in
the electromagnetic case and, for gravitons, it has the same
derivative structure as the Weyl tensor. 
Their associated  Lagrangians read ${\cal L}_I=\phi^2$
for the scalar field, ${\cal L}_I=F_{\mu\nu}F^{\mu\nu}$ for
photons and ${\cal L}_I=C_{\mu\nu\rho\sigma}C^{\mu\nu\rho\sigma}$
for gravitons, which coincide with the results obtained for
asymptotically flat wormholes.

 The interaction Lagrangians for
higher modes in adS spacetime and those in the flat case differ
in terms that are proportional to powers of the cosmological constant. 
For instance, the effective interaction Lagrangians coming from 
the $n=3$ inhomogeneous mode of a scalar field is of the form
$(\nabla^\mu\nabla^\nu\phi-\frac{1}{6}\Lambda g^{\mu\nu}\phi)^2$; 
similarly, we can easily find that 
the $n=4$ inhomogeneous mode of the electromagnetic field
provides an effective interaction Lagrangian of the form 
$(\nabla^\mu\nabla^\nu F^{\rho\sigma}-\frac{1}{3}\Lambda 
g^{\mu\nu}F^{\rho\sigma})^2$.
It
is also worth noting that adS wormholes do not induce any direct
modification to the cosmological term nor to Newton's constant.

\section*{Acknowledgments}

We are very grateful to Guillermo A. Mena Marug\'an, Mariano
Moles and Pedro F. Gonz\'alez-D\'{\i}az for helpful discussions
and suggestions. C.B was supported by a Spanish Ministry of
Education and Culture (MEC) grant. C.B. is also grateful to James
Hartle and the Institute for Theoretical Physics (UCSB), where
part of this work was done, for warm hospitality. This research
was supported in part by the National Science Foundation under
Grant No. PHY94--07194. L.J.G. was supported by funds provided by
DGICYT and MEC under Contract Adjunct to the Project No.
PB94--0107.

\end{document}